\documentclass[journal,10pt]{IEEEtran}

\normalsize

\ifCLASSINFOpdf
 \usepackage[pdftex]{graphicx}
\else
 \usepackage[dvips]{graphicx}
\fi

\DeclareGraphicsExtensions{.eps}
\usepackage[cmex10]{amsmath}
\usepackage{bm}
\usepackage{upgreek}
\usepackage{epsfig}
\usepackage[noadjust]{cite}
\usepackage{latexsym}
\usepackage{multirow}
\usepackage{epstopdf}
\usepackage{color}  
\usepackage{amssymb}
\graphicspath{{./Figures/}}
\usepackage{color}
\usepackage{bbm}
\usepackage{mathtools}
\usepackage{url}

\usepackage{cite}
\usepackage[tight,footnotesize]{subfigure}
\usepackage{balance}

\usepackage{color, soul}

\usepackage{siunitx}
\DeclareSIUnit\bps{bps}
\DeclareSIUnit\Torr{Torr}
\DeclareSIUnit\torr{Torr}
\DeclareSIUnit\sample{Sa}

\usepackage{cite}

\begin{document}

\title{THz ISAC: A Physical-Layer Perspective of Terahertz Integrated Sensing and Communication}
%
%
\author{Chong~Han, Yongzhi~Wu, Zhi~Chen, Yi~Chen, and Guangjian~Wang
}

\maketitle

\begin{abstract}
The Terahertz (0.1-10 THz) band holds enormous potential for supporting unprecedented data rates and millimeter-level accurate sensing thanks to its ultra-broad bandwidth. Terahertz integrated sensing and communication (ISAC) is viewed as a game-changing technology to realize connected intelligence in 6G and beyond systems. 
In this article, challenges from THz channel and transceiver perspectives, as well as difficulties of ISAC are elaborated.
Motivated by these challenges,
THz ISAC channels are studied in terms of channel types, measurement and models. Moreover, four key signal processing techniques to unleash the full potential of THz ISAC are investigated, namely, waveform design, receiver processing, narrowbeam management, and localization. Quantitative studies demonstrate the benefits and performance of the state-of-the-art signal processing methods. Finally, open problems and potential solutions are discussed.
\end{abstract}


%
\IEEEpeerreviewmaketitle

\section{Introduction}
%
%
%
%
\IEEEPARstart{W}{ith} the revolutionary enhancement of wireless data rates and the increasing demand of highly accurate sensing capability, a promising blueprint is expected to be realized: all things are sensing, connected and intelligent~\cite{Tong20216G}. The Terahertz (THz) band, and integrated sensing and communication are two descriptive technologies for achieving such a vision in sixth generation (6G) and beyond wireless systems, by exploring the new spectrum as well as the new function of sensing, respectively. Two forces drive together to motivate the study of integration of these two technologies. On one hand, wireless links with Terabit-per-second (Tbps) will come true in the future intelligent information society. On the other hand, the ultra-broad bandwidth in the band also provides huge potential to realize ultra-accurate sensing, e.g., millimeter-level and millidegree-level accuracy.

\begin{figure*}
    \centering
    \includegraphics[width=0.88\textwidth]{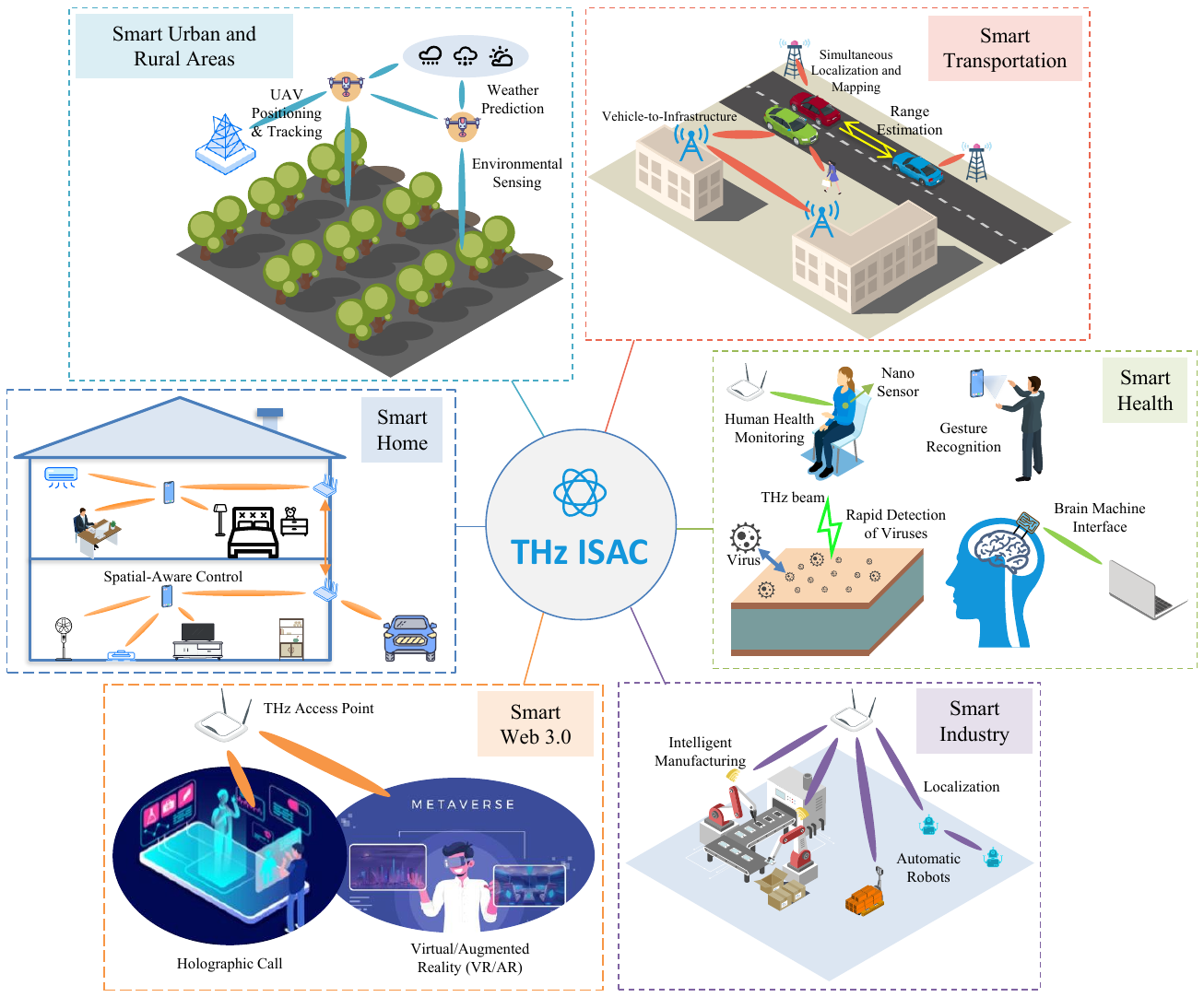}
    \caption{Applications driven by THz ISAC, including: (a) smart urban and rural areas, (b) smart transportation, (c) smart home, (d) smart health, (e) smart web 3.0,  (f) smart industry.}
    \label{fig:THz_ISAC_applications}
\end{figure*}

Thanks to the abundant continuous bandwidth in the THz band~\cite{chen2021THz}, the integration of THz sensing and communication (THz ISAC) can enable simultaneously transmitting billions of
data streams and ultra-accurate sensing by sharing the spectrum, hardware, and signal processing modules. In addition to enhance the performance and functionality, THz ISAC is able to effectively reduce the hardware cost, and improve the
spectral and energy efficiency~\cite{liu2022isac}. In light of these, THz ISAC is envisioned to provide high quality-of-experience (QoE) for various future services and applications demanding for ultra-high data rates and ultra-high resolution sensing. As shown in Fig.~\ref{fig:THz_ISAC_applications}, we envision the applications driven by THz ISAC technologies including smart urban and rural areas, smart transportation, smart home, smart health, smart web 3.0, and smart industry, among many other emerging ones to appear.

With an increasing level of integration, THz ISAC systems can be divided into the following four categories~\cite{zhang2022jcs}. 
\begin{itemize}
    \item At the bottom level, communication and sensing coexistence improves the spectral efficiency by encouraging frequency reuse for these two functionalities. 
    \item At the second level, transmitting separate signals with partially or fully shared system hardware is helpful to reduce the cost, size and weight of the system.
\item To reach a higher level of integration, signal processing can be shared to jointly design and optimize the transmit waveform used for both communication and sensing.
\item With full-fledged integration, THz ISAC systems evolve more successfully when the protocol and network design is shared beyond the physical layer design.
\end{itemize} 
In this article, we investigate channel modeling and signal processing of THz ISAC from the physical layer perspective. First, we analyze the challenges of THz ISAC and difficulties of the integration, due to the peculiarities of THz channels and transceivers. Next, we analyze four signal processing techniques for THz ISAC systems, including transmit waveform, receiver processing, narrowbeam management and localization. For each of them, problems and possible solutions are delineated respectively. To assess the performance, numerical evaluations of THz ISAC channel, waveform and receiver signal processing methods are elaborated. Finally, some open problems and potential research directions of THz ISAC are presented.

\section{Challenges of THz ISAC}

Despite the wonderful vision of the THz spectrum, challenges are still encountered from two perspectives of THz propagation and devices. On one hand, the signal processing methods for THz ISAC, including the transmitter and receiver design, should be tailored for specific THz communication and sensing channels to make the best of the available spectrum. On the other hand, since the transceiver impairments increase drastically at higher frequencies, the signal processing strategy is more affected by the transceiver features when it comes to the THz band. Furthermore, challenges on integrating the sensing function with communication are presented.

\subsection{Challenges from THz Channel Perspectives}

\subsubsection{Propagation Delay}
Since the reflection and scattering losses at THz frequencies depend on the angle of incidence and may result in strong power loss of a non-line-of-sight (NLoS) path and the decrease of the number of the dominant rays, the delay spread of THz communication channel is reduced. Since cyclic prefix (CP) duration is usually set longer than the maximum delay spread, the CP length in the THz band can be reduced to improve the spectral efficiency. Nevertheless, for classical orthogonal frequency division multiplexing (OFDM) sensing systems, the round-trip delay of the maximum sensing distance should be smaller than CP duration. With the decrease of communication delay spread, the CP duration is reduced in THz ISAC systems and limits the sensing distance even when the link budget is sufficient~\cite{wu2022flexible}.

\subsubsection{Doppler Effects}

Being proportional to the carrier frequency, the Doppler spread effect becomes even stricter in the THz band, especially in high-mobility scenarios. Doubly-selective channels with high Doppler spreads may breaks orthogonality of subcarriers and cause inter-carrier interference.
Increasing the subcarrier spacing can mitigate the Doppler effect, but it may sacrifice the velocity estimation accuracy.

\subsubsection{Blockage Problem}

Due to the strong penetration loss,
the blockage effect needs to be considered in THz ISAC. THz waves nearly lose the ability to go through the common blockages, such as walls and human bodies. In an environment encountering various blockages, the line-of-sight (LoS) path between an access point and a user equipment might be blocked and thus, the coverage distance of the access point would be reduced. Meanwhile, blockages may affect the accuracy of THz localization and map construction in realistic environments, since they can block some THz rays and reduce number of channel features.

\subsubsection{Near- and Far-field Propagation}

In the THz band, when transmission distance is shorter than the Rayleigh distance of the antenna array, we need to consider near-field propagation. In this case, the planar-wave propagation, which is an approximation of spherical-wave propagation, becomes invalid and causes difficulties in beamforming. Moreover, the sensing target in the far-field of the transmitter can be considered as a point target, while an extended or surface model is required for targets in the near-field.

\subsection{Challenges from THz Transceiver Perspectives}
\label{sec:transceiver_challenge}
\subsubsection{Power Amplifier Efficiency}

Noticeably, the saturated output power of power amplifiers (PAs) rapidly decreases as the carrier frequency increases. Thus, the PA efficiency and average output power of transmitters are more sensitive to the peak-to-average power ratio (PAPR) in the THz band. In order to optimize the transmit power and power efficiency of the transmitter, the key step is to reduce the PAPR of THz ISAC transmit signal.

\subsubsection{Beam Squint Effect}

In the THz band, the ultra-large dimensional antenna arrays (e.g., over 1000~\cite{han2021beamforming}) are leveraged to provide high beamforming gain. Herein, a beam squint effect arises in wideband ultra-masssive multiple-input multiple-output (UM-MIMO) with analog or hybrid beamforming. Without full freedom of beam weighting, hybrid or passive structures utilize phase shifters that tune the same weights for all frequencies, and thus the beam directions may deviate away from the central one and off the target. As a result, the array gain is substantially reduced, which affects the performance of sensing and communication.

\subsubsection{Implementation Complexity}

To support wireless links with ultra-fast data rates, THz transceivers are required to be well designed in terms of the implementation complexity. Since complex transceivers with Tbps digital processors are still challenging, low-complexity signal processing methods are preferred, especially receiver processing, including channel parameter estimation and data detection. The algorithms should be designed with logarithmic or linear time complexity.

\subsection{Challenges of Integration of Sensing and Communication}
Integrating sensing function with communication encounters several challenges due to different requirements and objectives of signal processing and separate channels. First, when sensing and communication are scheduled on non-overlapped wireless resources in time, frequency, spatial and code domains, resource allocation schemes need to be coordinated and optimized~\cite{liu2022isac}. Second, if sensing and communication are integrated with a unified ISAC waveform, these two functionalities have different requirements on the waveform and modulations. Specifically, sensing demands good autocorrelation properties to realize high resolution, while communication prefers high efficiency and reliability. Third, the separation and intersection of sensing and communication channels are influenced by types of sensing services, e.g., active or passive sensing, which apply different designs for sensing receivers with or without knowledge of transmit signal. Fourth, since sensing aims at estimating target parameters while communication needs to recover transmit data, signal processing for sensing and communication is generally hard to be integrated, which motivates novel multi-task receiver design.

In this section, the influence of some unique features on THz ISAC are described, such as propagation delay and spherical wave propagation. In addition, compared with the millimeter-wave (mmWave) band, many effects become severer and cause stricter requirement on signal processing of THz ISAC, e.g., stronger Doppler effect, decreased saturated output power of PAs, and worsening beam squint effect. Considering these, innovative studies are presented in terms of THz ISAC channel, transmitter and receiver signal processing.

\begin{table*}[]
\centering
\caption{THz ISAC Waveform Comparison}\label{tab:waveform}
\includegraphics[width=0.9\textwidth]{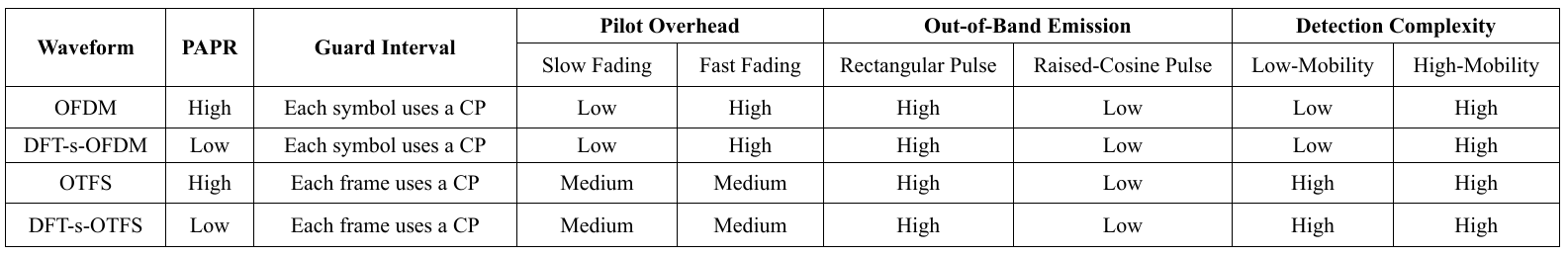}
\end{table*}

\section{THz ISAC Channel}
\subsection{ISAC Channel Types}
A communication channel is medium that transports information. By contrast, in the perspective of sensing, a channel is the information itself. In particular, the context of distance, dimension, shape, velocity and other properties of the target and the environment are self-contained within the power, delay, Doppler shift and angular information of the multipaths in a channel. According to the deployment of the transmitter (Tx) and receiver (Rx), the ISAC channel can be summarized into three types, namely, monostatic, bistatic and distributed channels.
\subsubsection{Monostatic ISAC Channel}
The transmitter and receiver of an monostatic ISAC channel are co-located. Monostatic channel can sense the target by the echoes of the transmitted THz signal, which contains the distance information of the target. To reconstruct the three-dimensional (3D) environment, the monostatic ISAC transceivers scan in the both azimuth and elevation angle domains. However, a monostatic ISAC channel consists of strong self-interference signal leaked by the transmitter, which needs to be carefully eliminated. In addition, multipath propagation may lead to low signal-to-noise-ratio (SNR) values of the desired echoes.
\subsubsection{Bistatic ISAC Channel}
A bistatic ISAC channel is simply like classic communication channels, where transmitter and receiver are physically separated from each other. Therefore, self-interference is naturally avoided in bistatic ISAC channels. Compared with a monostatic ISAC channel, multipath rays in a bistatic ISAC channel offer rich power, distance and angle information of the sensing target, which interacts with the THz signal. Nevertheless, stringent synchronization between Tx and Rx  is required, without which sensing performance degradation could happen due to the phase instability of the sensing signal. The information of the scatterers in the space are extracted and estimated through the post-processing techniques. Considering spherical wave in the post-processing is critical for improving sensing accuracy.
\subsubsection{Distributed ISAC Channel}
There are more than one transmitter or receiver in a distributed ISAC channel. The distributed ISAC channels provide spatial diversity of objects, which are helpful to improve sensing performance. However, interference from the transmitters in a distributed ISAC system should be mitigated with careful coordination and proper cancellation techniques. At the receiver, data fusion strategies are employed to make full use of the multiple copies of information, including complementary, redundant, and cooperative replicas contained in the distributed ISAC channel.

\subsection{Channel Measurement and Modeling}
Channel measurement is to obtain the channel response of the channel between the transmitter and receiver. The channel response contains the information about the gain and delay of the multipaths. The angular information of the multipaths draws much interest in ISAC channels. Therefore, the focus of channel measurement extends from the time-of-arrival (ToA) to the angle of arrival and angle of departure. There are three categories of channel sounding methods for the wideband channel measurement.
First, a vector network analyzer (VNA)-based frequency-domain channel sounder directly measures the frequency response of the channel. We have developed VNA-based THz ISAC channel measurement systems at 100-400 GHz~\cite{Han2022COMST}.
Second, a correlation-based sounding method directly measures the channel impulse response from the received sequence, based on sliding correlation. Third, unique in the THz band, a THz time-domain spectroscopy (THz-TDS) can be used for sample detection and imaging due to its wide spectrum up to several THz.

On the basis of whether invoking statistical distributions in a model, we categorize the modeling approaches of THz ISAC channels as deterministic, statistical, and hybrid methods~\cite{Han2022COMST}. In comparison, deterministic channel models are more suitable for deployment planning, while statistical channel models are useful for development and testing of wireless systems.
To combine the benefits from different modeling methods, a hybrid channel models can achieve good trade-off between accuracy and complexity. Specifically in the THz ISAC channels, ray-tracing can capture the echoes of the target for sensing, while statistical parts are adopted to describe the multipath propagation for communication~\cite{chen2021channel}.

\section{Signal Processing for THz ISAC}

\begin{table*}[]
\centering
\caption{Receiver Signal Processing Methods for THz ISAC Waveform}\label{tab:receiver}
\includegraphics[width=0.88\textwidth]{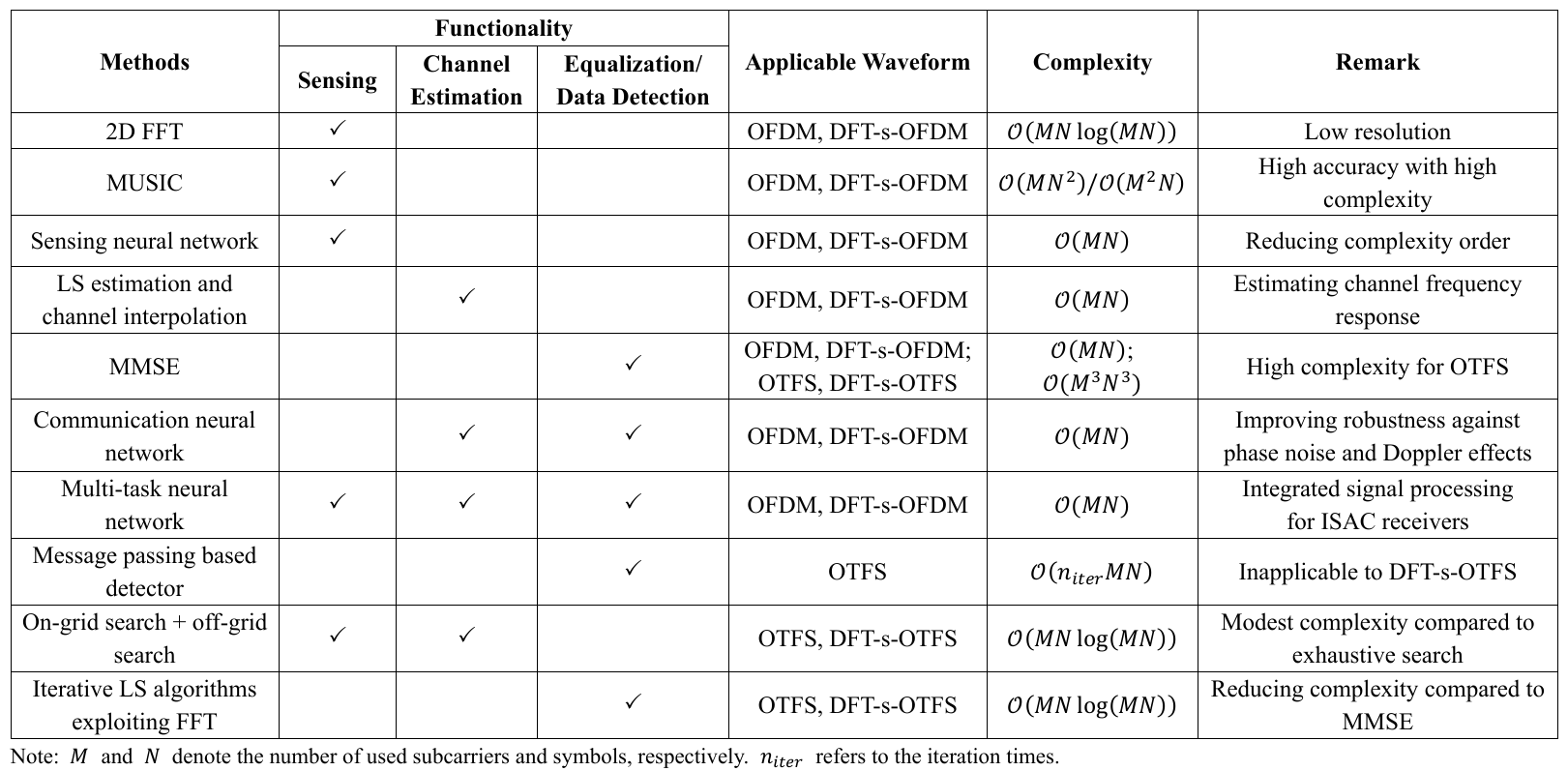}
\end{table*}

\subsection{Waveform Design}

Sensing and communication can fully share the hardware and signal processing modules when jointly design the ISAC transmit waveform, which thereby reduce power consumption and signal processing complexity. OFDM is a potential ISAC waveform, since it has good compatibility with 4G and 5G standards. However, OFDM waveform  inherently suffers from a high PAPR and might induce undesirable clipping distortions in THz PAs~\cite{mao2022waveform}. Meanwhile, the Doppler shifts are difficult to handle in the time-frequency domain, while Doppler effects become stronger at THz frequencies.

When it comes to the single-carrier counterpart, DFT-s-OFDM and its variants are regarded as more potential candidate waveforms for THz ISAC, thanks to their low PAPR compared with multi-carrier waveforms.
DFT-s-OFDM data signals present Gaussian randomness in the frequency domain, which causes severe noise amplification in point-wise division of OFDM sensing~\cite{wu2022flexible}. Thus, constant-enveloped pilot signals can be employed to conduct sensing parameter estimation.

By contrast, orthogonal time frequency space (OTFS) modulation can deal with Doppler effects and conveniently accommodate the channel dynamics in the delay-Doppler domain~\cite{wei2021otfs}. Nevertheless, the PAPR of OTFS is still not satisfactory considering THz PAs and the receiver processing complexity of OTFS remains a pivotal issue.
Recently, a DFT spread OTFS (DFT-s-OTFS) waveform is proposed in~\cite{wu2022dftsotfs} to reduce the PAPR of OTFS, by developing a DFT precoding operation on the information symbols along the Doppler axis.
In Table~\ref{tab:waveform}, the aforementioned THz ISAC waveforms are compared in terms of the PAPR, guard interval design, pilot overhead, out-of-band emission, and detection complexity.

\subsection{Receiver Signal Processing}
\label{sec:rec_sig}

Despite the promising sensing and communication abilities of THz ISAC waveforms, it is crucial to design signal processing methods at the receiver side, due to the following challenges. 
For OFDM and DFT-s-OFDM systems, the link performance might be seriously deteriorated by severe non-linear distortion effects at the THz transceivers, such as phase noise effects and Doppler effects in high-mobility scenarios (as discussed in Sec.~\ref{sec:transceiver_challenge}). In this case, deep learning based methods are developed to improve the sensing and communication performance, thanks to the stronger robustness against non-ideal conditions in contrast with conventional methods.

While OTFS is proposed to deal with the Doppler effects, the implementation complexity of OTFS receiver is a significant challenge. To be specific, channel parameter estimation and data detection of OTFS have high complexity. To realize high-speed baseband digital processing in the THz band, the computational efficiency is of great concern. Message passing and Bayesian learning based methods with reduced complexity are developed to perform channel estimation and data detection in OTFS systems~\cite{wei2021otfs}. Nevertheless, if the information symbols are not directly mapped on the delay-Doppler domain, these approaches become inapplicable. A channel equalizer for DFT-s-OTFS systems by using iterative least squares (LS) algorithms is designed in~\cite{wu2022dftsotfs}, since the derived channel matrix with fractional delay and Doppler has partial Fourier matrices and the FFT algorithm can be employed to reduce the complexity.
In Table~\ref{tab:receiver}, the receiver signal processing methods are compared in terms of functionality, applicable waveform and complexity. While signal processing techniques are presented for exploring the atomic behavior of materials in THz sensing~\cite{sara2022thz} and learning the interaction between THz sensing, imaging and localization~\cite{2020THzSensing}, we investigate signal processing methods in terms of sensing, channel estimation and data detection with the aforementioned potential candidate waveforms for THz ISAC.

\subsection{Narrowbeam Management}\label{sec:beam}

When using directional beams to overcome severe path loss in the THz band, communication and sensing have different objectives, which may cause conflict on the beam directions. On one hand, sensing prefers scanning beams in search phase and accurate beams towards targets in estimation phase. On the other hand, communication requires several stable beams towards multiple users. When THz channels are shared between sensing and communication, the angle-of-departure (AoD) and angle-of-arrival (AoA) information of communication beams can be exploited to assist sensing and localization in beam-based THz ISAC systems. Thus, effective narrowbeam management schemes are important to realize ISAC in THz UM-MIMO systems, including beamforming design, beam scanning, beam tracking and cooperative beamforming.

To manage multiple narrowbeams for THz ISAC systems, two stages including target searching and target estimation can be established. At the first stage, scanning beams are generated to search possible targets in the surrounding environment. The beam squint effect causes the beam misalignment from the central direction at non-central frequencies, which can be exploited to widen the scanning sensing beams and accelerate seeking for a target of interest. At the second stage, when estimating parameters of the sensing target, the beam squint effect needs to be mitigated to focus the sensing beam on the target direction. A promising solution to solve this problem is to use THz dynamic-subarray with fixed-true-time-delay (DS-FTTD) hybrid beamforming architecture~\cite{han2021beamforming}.
For array-of-subarrays (AoSA) or fully-connected hybrid beamforming structures, the subspace-based methods, including MUSIC and ESPRIT, can achieve super-resolution angle estimation by exploiting the subspace conducted from eigenvalue decomposition. To further address the problems of super precision AoA estimation and tracking in THz dynamic array-of-subarrays (DAoSA) systems, the DL-based methods can be designed to capture the angle variation with reduced computational complexity and beamforming training overhead.

\subsection{Localization and Map Reconstruction}

As one of popular sensing services, high-accuracy localization and map reconstruction have gained increasing attention. Thanks to the ultra-broad bandwidth, centimeter-level and even millimeter-level localization can be realized by employing the channel state information (CSI) of THz signals, including AoA, received power, and propagation delay~\cite{fan2021localization}. While the geometry-based and learning-based algorithms for THz localization systems are investigated in~\cite{chen2022localization} based on these input features, we can use the aforementioned receiver processing methods to extract them from the THz ISAC signals, as elaborated in Sec.~\ref{sec:rec_sig} and~\ref{sec:beam}.

It is a significant problem to realize high-precision localization based on the above CSI. A DL-based solution can be developed to solve the 3D THz indoor localization problem with an offline stage and an online stage~\cite{fan2021localization}. At the offline stage, based on multiple THz access points (APs), we first extract the CSI features from the received signals at the user equipment (UE) to build the dataset and train a neural network model.
Then at the online stage, we deploy the trained DL model at the UE, which is employed to predict 3D UE coordinates and realize ultra-accurate real-time localization.

\section{Performance Evaluations}

In this section, we provide numerical results to quantitatively illustrate the channel properties and the performance of the state-of-the-art signal processing methods for THz ISAC.

\subsection{THz ISAC Channel}
\begin{figure}
    \centering
    \includegraphics[width=0.4\textwidth]{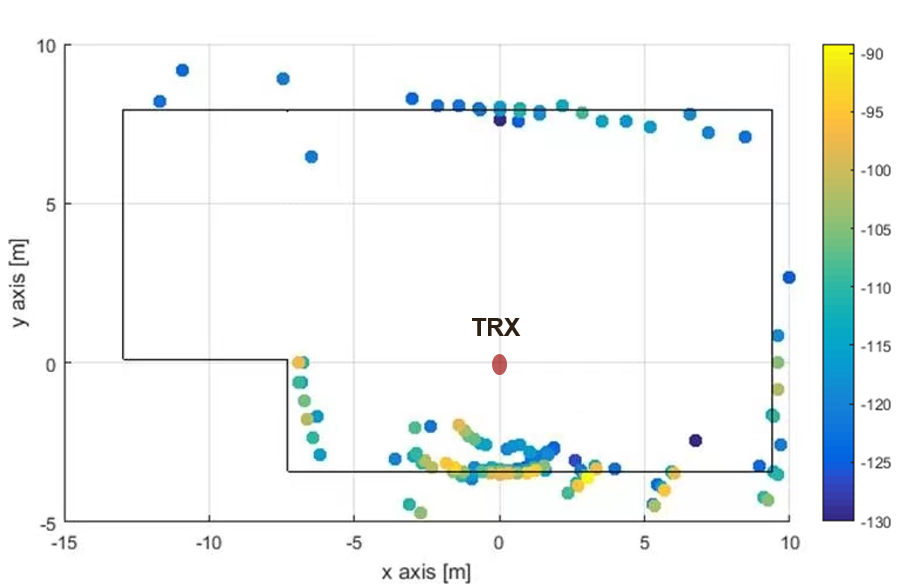}
    \caption{Monostatic ISAC channel measurement in an office room at 140~GHz.}
    \label{fig:monostatic-office}
\end{figure}

Monostatic and bistatic channel measurement in the low THz band has been conducted in indoor meeting room, office room and outdoor scenarios. 
In particular, a monostatic channel measurement campaign at 140 GHz from~\cite{chen2021channel} has been conducted for sensing an office room. As shown in Fig.~\ref{fig:monostatic-office}, the transmitter/receiver (TRX) is surrounded by cubicles and the power of the received THz echoes from the scanned directions are represented by different colors. The transmitter scans the surrounding environment in the azimuth domain from $0^{\circ}$ to $360^{\circ}$ with a spatial rotation step of $1^{\circ}$. The receiver extracts the round-trip delay from the channel impulse response and identifies the locations of the cubicles and the walls of the office room.
To assess the accuracy of reconstruction, the error between the reconstructed geometry and the measured office room is calculated to follow an exponential distribution, with the mean value of 0.39 meters.

\subsection{Signal Processing for THz ISAC}

\begin{figure}
    \centering
    \includegraphics[width=0.4\textwidth]{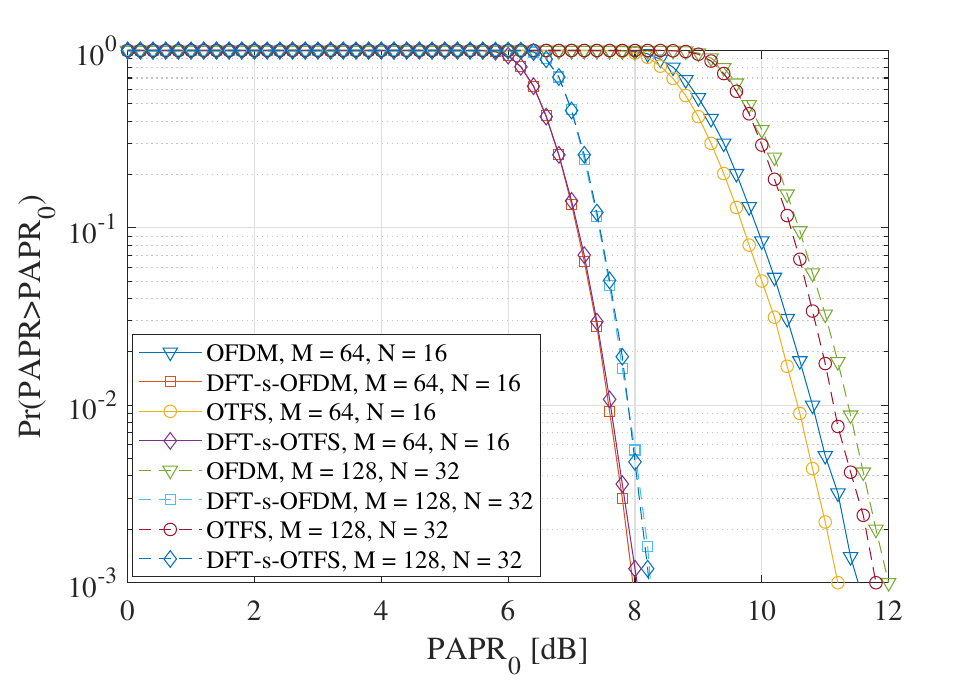}
    \caption{PAPR comparison for transmit signal of OFDM, DFT-s-OFDM, OTFS and DFT-s-OTFS with respect to different number of subcarriers $M$ and symbols $N$.}
    \label{fig:PAPR}
\end{figure}

\begin{figure}
    \centering
    \includegraphics[width=0.4\textwidth]{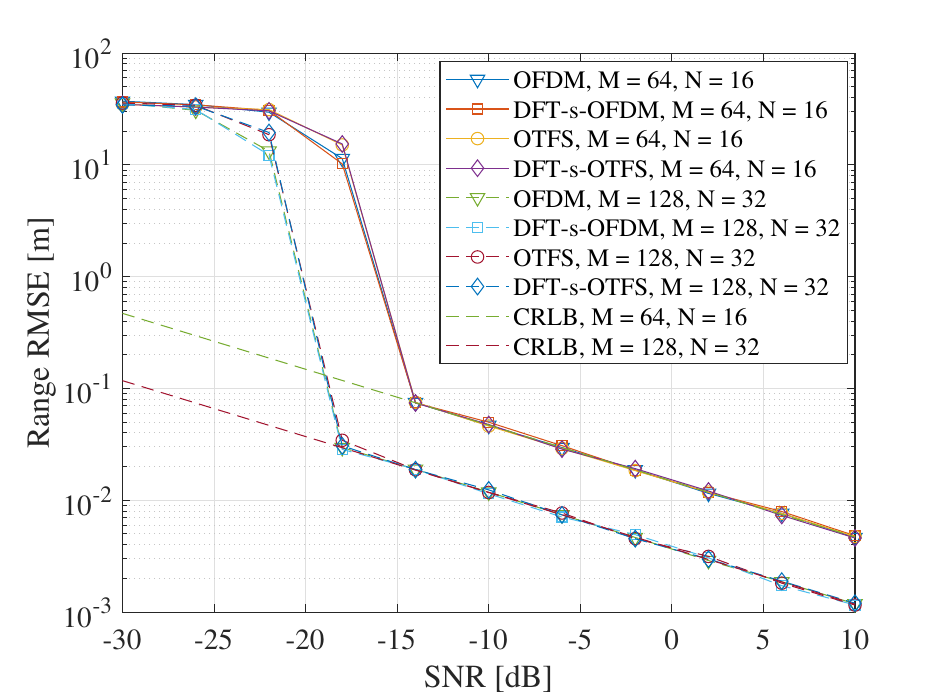}
    \caption{The root mean square errors (RMSEs) of target range estimation are compared among various waveforms with respect to different number of subcarriers $M$ and symbols $N$. The RMSE can achieve the Cram\'er-Rao lower bound (CRLB) when the SNR is above -15 dB.}
    \label{fig:waveform_range_rmse}
\end{figure}

While comparisons between OFDM and OTFS have been studied in the literature, we introduce another two candidate waveforms for THz ISAC, i.e., DFT-s-OFDM and DFT-s-OTFS, to compare their PAPR performance and sensing accuracy. To list the key simulation parameters, the carrier frequency equals to 0.3 THz. The subcarrier spacing is set as 1.92 MHz. 4-QAM modulation scheme is employed. The PAPR of these waveforms is evaluated by conducting 4-times oversampling on the discrete-time baseband signal.
In Fig.~\ref{fig:PAPR}, we demonstrate that the PAPR of DFT-s-OTFS approaches DFT-s-OFDM, while their PAPRs are approximately 3 dB lower than OTFS and OFDM. 

Next, we set a sensing target with the distance of 10 m and the relative velocity of 20 km/h.
For sensing parameter estimation, we use the 2D DFT algorithm for OFDM and DFT-s-OFDM~\cite{zhang2022jcs}, while employing the proposed two-stage estimation method in~\cite{wu2022dftsotfs} for OTFS and DFT-s-OTFS, i.e., on-grid search to estimate the integer tap of delay and Doppler, and off-grid estimation with the FFT algorithm.
As shown in Fig.~\ref{fig:waveform_range_rmse}, we learn that all of these waveforms can achieve millimeter-level range estimation accuracy. Moreover, the range estimation error can be reduced by increasing the subcarrier number from 64 to 128 and the symbol number from 16 to 32.

In summary, in low-mobility channels, OFDM is still an excellent candidate for THz ISAC, and achieves good sensing and communication performance, since it enables good compatibility with UM-MIMO, flexible multi-user scheduling and resource allocation. In high-mobility scenarios, DFT-s-OTFS provides stronger robustness to severe Doppler spread by exploiting the channel sparsity of delay-Doppler domain, while it comes at the price of increased detection complexity. For energy-constrained links, DFT-s-OFDM and DFT-s-OTFS are shown to realize high energy efficiency due to their low PAPR characteristics.

\section{Open Problems and Potential Research Directions}

\subsection{THz ISAC Transmission}
Despite the PAPR issue, OFDM is still a potential candidate waveform for THz ISAC. Without inter-symbol interference, OFDM enables flexible multiple access, i.e., allowing multiple users to select their own subset of time-frequency resource blocks. In addition, sensing-centric waveform with embedded information data may improve the sensing resolution while sacrificing the spectral efficiency.

Besides the transmit waveform, for mono-static sensing scenarios, we need to consider the duplex schemes of THz ISAC systems, including half-duplex, full-duplex or hybrid-duplex (i.e., communication is half-duplex while radar is full-duplex). When using half-duplex schemes, the sensing distance is limited by the switching time between transmitter and receiver mode at both the base station (BS) and UE. In addition, it is still an open issue whether the functionality of THz sensing should be supported at BS, UE or both of them.

\subsection{THz ISAC with UM-MIMO}

While THz ISAC systems are usually enabled with UM-MIMO, the compatibility between UM-MIMO and transmit waveform needs to be considered. Moreover, with narrow beams generated by UM-MIMO, interference management is still a significant issue, such as intra-cell and inter-cell interference management when performing THz sensing. In this case, unique sensing sequences in UM-MIMO systems are required to be interference-resistant, in which their length and amount need to be optimized for different scenarios, including vehicular networks, outdoor and indoor wearable/handheld applications.

In THz ISAC systems with UM-MIMO, sharing of spatial resources is indispensable. To some extent, the beamforming gain of wireless links needs to be controlled, since the energy might leak in other angles apart from the direction toward the desired user. Thus, to achieve a satisfying trade-off between high communication capacity and high-resolution sensing ability, spatial resources are required to be flexibly scheduled while satisfying the sensing requirements.

\subsection{Immersive Resource Allocation}

Besides the spatial resource allocation, frequency-division, time-division and mixed schemes for resource allocation among communication and sensing need to be investigated. In a communication frame, several types of signals can be employed for sensing, such as data payload signals, signal synchronization block (SSB), and demodulation reference signals (DMRS). However, it is still not determined which type and how much resource of these signals for THz sensing. Moreover, sharing such resources results in performance tradeoff between sensing accuracy and communication capacity.

In addition to the above allocation problems, it is an open issue to intelligently allocate the associated access points for users and sensing functionality in presence of blockages. Artificial intelligence (AI) techniques can be leveraged to optimize the resource allocation schemes in such scenarios.

\section{Conclusion}

THz ISAC is one of key promising enablers to meet demands for next-generation wireless systems. In this paper, channel modeling and signal processing techniques are investigated to overcome the challenges and improve the performance of THz ISAC. Channel modeling and measurement methods of THz ISAC are summarized. Four signal processing technologies, including waveform design, receiver signal processing, narrowbeam management and localization, are thoroughly analyzed and evaluated.
Finally, open problems and potential research directions are stated.


%



\ifCLASSOPTIONcaptionsoff
  \newpage
\fi



%

\bibliography{main}
\bibliographystyle{IEEEtran}

%

%


\balance
%
%



\section*{Acknowledgement}
The work of Chong Han was supported in part by the National Key Research and Development Program of China under Project 2020YFB1805700.

\section*{Biographies}
\vspace{-1cm}
\begin{IEEEbiographynophoto}{Chong Han} is John Wu \& Jane Sun Endowed Associate Professor with the Terahertz Wireless Communications (TWC) Laboratory, Shanghai Jiao Tong University, China. 
\end{IEEEbiographynophoto}
\vspace{-1cm}
\begin{IEEEbiographynophoto}{Yongzhi Wu} is currently pursuing a Ph.D. degree at the Terahertz Wireless Communications (TWC) Laboratory, 
Shanghai Jiao Tong University, China.
\end{IEEEbiographynophoto}
\vspace{-1cm}
\begin{IEEEbiographynophoto}{Zhi Chen} is a Professor with University of Electronic Science and Technology of China.
\end{IEEEbiographynophoto}
\vspace{-1cm}
\begin{IEEEbiographynophoto}{Yi Chen} received his Ph.D degree from Shanghai Jiao Tong University, China in 2022 and joined Huawei since then.
\end{IEEEbiographynophoto}
\vspace{-1cm}
\begin{IEEEbiographynophoto}{Guangjian Wang} is a senior research scientist with Huawei Technologies, China.
\end{IEEEbiographynophoto}

\balance
\end{document}